\documentclass[letterpaper, 10 pt, conference]{ieeeconf}

\IEEEoverridecommandlockouts 
\usepackage{cite}
\usepackage{amsmath,amssymb,amsfonts}
\usepackage{url}
\usepackage{color}
\usepackage{hyperref}
\usepackage{graphicx}
\usepackage{lipsum}  
\usepackage{optidef}

\usepackage{apptools}
\AtAppendix{\counterwithin{theorem}{subsection}}
 
\newtheorem{theorem}{Theorem}[section]
\newtheorem{lemma}[theorem]{Lemma}

\newtheorem{remark}{Remark}
\newtheorem{assumption}[theorem]{Assumption}
\newtheorem{example}{Example}

\newcommand{\until}[1]{\{1,\dots,#1\}}
\newcommand{\fromto}[2]{\{#1,\dots,#2\}}

\newcommand{\vect}{\operatorname{vec}}
\newcommand{\Ker}{\operatorname{Ker}}

\newcommand{\Rank}{\operatorname{Rank}}
\newcommand{\Image}{\operatorname{Im}}

\newcommand{\real}{\mathbb{R}}
\newcommand{\complex}{\mathbb{C}}

\newcommand{\Basis}{\operatorname{Basis}}

\newcommand{\mc}{\mathcal}

\newcommand\ubf{\mathbf{u}}

\newcommand\xbf{\mathbf{x}}

\newcommand{\norm}[1]{\left\lVert#1\right\rVert}

\newcommand\oprocendsymbol{\hbox{$\square$}}
\newcommand\oprocend{\relax\ifmmode\else\unskip\hfill\fi\oprocendsymbol}

\newcommand*{\QEDA}{\hfill\ensuremath{\blacksquare}}

\graphicspath{{../../img/}}

\begin{document}
\title{\LARGE \bf Data-driven Eigenstructure Assignment for Sparse
  Feedback Design}

\author{Federico Celi, Giacomo Baggio, and Fabio Pasqualetti
  \thanks{This material is based upon work supported in part by awards
    ARO W911NF-20-2-0267, AFOSR-FA9550-20-1-0140, and
    AFOSR-FA9550-19-1-0235. Federico Celi and Fabio Pasqualetti are
    with the Department of Mechanical Engineering, University of
    California at Riverside,
    \{\href{mailto:fceli@engr.ucr.edu}{\texttt{fceli}},
    \href{mailto:fabiopas@engr.ucr.edu}{\texttt{fabiopas\}@engr.ucr.edu.}}
    Giacomo Baggio is with the Department of Information Engineering,
    University of Padova, Italy
    \href{mailto:baggio@dei.unipd.it}{\texttt{baggio@dei.unipd.it}}.}}
\maketitle


\begin{abstract} 
This paper presents a novel approach for solving the pole placement and eigenstructure assignment problems through data-driven methods. By using open-loop data alone, the paper shows that it is possible to characterize the allowable eigenvector subspaces, as well as the set of feedback gains that solve the pole placement problem. Additionally, the paper proposes a closed-form expression for the feedback gain that solves the eigenstructure assignment problem. Finally, the paper discusses a series of optimization problems aimed at finding sparse feedback gains for the pole placement problem. 
\end{abstract}

\section{Introduction}
Data-driven control methods have gained popularity in recent years thanks to their ability to synthesize feedback controllers directly from historical data \cite{GB-DSB-FP:20,MP-SS-MT-CC-AK-RS-JN:18,XX:22}. One major advantage of these methods is that they eliminate the need for constructing or identifying a model for the underlying system to be controlled. This is especially useful in situations where deriving first-principle models is challenging or where the identification process may lead to unreliable model parameters. 
Despite the popularity of data-driven control, the problems of data-driven pole placement and eigenstructure assignment have not been explored until recently \cite{SM-RRH:22,GB:23}. 

The traditional (i.e., model-based, non-sparse) pole placement and eigenstructure assignment problems have a rich history, including in practical applications \cite{MI-CS:90,ANA-EYS-JCC:83}. The pole placement (eigenstructure assignment) problem consists in finding a static feedback gain that produces a closed-loop system where the state matrix has a pre-specified set of eigenvectors (eigenvalues and eigenvectors) \cite{GPL-RJP:98}. We refer to the seminal works \cite{WMW:67,WMW-MAS:70,BCM:76,GK-BCM:77,JK-NKN-PVD:85} and to the recent papers \cite{FP-AF-LN:21,JQT-MGT-NSF-SJE:22}, which highlight the ongoing interest in these topics.

In general, the feedback gain which solves the pole placement problem is not unique, adding a certain degree of freedom on the choice of the feedback gain. This can be leveraged to enforce further control objectives, for example, by imposing a sparsity pattern on the feedback gain itself. By using a feedback with predefined sparsity patterns, or by maximizing the overall number of zero entries of the feedback gain, the number of feedback signals can be reduced while still achieving the desired closed-loop behavior. This can be advantageous in applications where the number of sensors or feedback signals is limited as, for example, in complex network systems \cite{FL-MF-MRJ:11al}.

In this paper, we address these problems and show that it is possible to place the closed-loop eigenvalues exactly at any desired location by designing a static feedback gain through open-loop data alone, i.e., without explicit knowledge of the system matrices. Further, we show that the static gain for eigenstructure assignment can be found through a closed-from data-driven expression. Finally, we apply these results to the design of sparse feedback gains.

\smallskip
\noindent
\textbf{Related work.}
Despite the recent advancements in data-driven control theory \cite{GB-VK-FP:19,JC-JL-FD:18,CDP-PT:19}, to the best of our knowledge, the only works discussing data-driven strategies for pole placement and eigenstructure assignment are \cite{SM-RRH:22} and \cite{GB:23} (the latter was developed concurrently and independently to the present paper). 
Both \cite{SM-RRH:22,GB:23} are based on the behavioral approach \cite{JWP-JCW:97} and rely on the \emph{Fundamental Lemma} \cite{JCW-PR-IM-BLMDM:05} to characterize the behavior of a linear system from a single, long, experimental trajectory. In contrast, our approach collects data from multiple trajectories, which has proven advantageous, e.g., when dealing with unstable systems, since shorter trajectories can be leveraged. A detailed analysis of the benefits of using multiple (shorter) trajectories over a single trajectory in control and reinforcement learning can be found in \cite{ST-RF-MS:22}.
It is worth noting that \cite{SM-RRH:22} uses Linear Matrix Inequalities to solve the problem of pole placement and it does not offer any closed-form solution, while \cite{GB:23} does not discuss a characterization of the set of static feedbacks for pole placement. Additionally, neither \cite{SM-RRH:22} nor \cite{GB:23} provide any insights on designing sparse feedback gains, which is a relatively unexplored topic even in the model-based framework \cite{FL-MF-MRJ:11al,VK-FP:18a}. This knowledge gap further motivates our interest in this problem. In \cite{JE-JC:22}, the authors propose a data-driven approach to designing sparse stabilizing feedback gains, however this method does not assign specific eigenvalues/eigenvectors as we do in this paper.
Recently, the System Level Approach to Controller Synthesis has proposed a set of tools for designing constrained robust, sparse, and optimal controllers, see \cite{YSW-NM-JCD:19}. However, the System Level Approach is based on designing a dynamic compensator, while the problems of pole placement and eigenstructure assignment are based on static feedback gains \cite{GPL-RJP:98}.


\smallskip
\noindent
\textbf{Paper contribution.}
This paper presents novel results on pole placement and eigenstructure assignment with sparse feedback achieved by using (possibly direct) data-driven formulas. Specifically, we characterize (i) the allowable eigenvector subspace and (ii) the set of feedback gains which solve the pole placement problem, both as a function of data. From these, we derive (iii) a closed-form expression of the gain that solves the eigenstructure assignment problem. Additionally, we (iv) discuss strategies for computing sparse feedback controllers for the pole placement problem, by incorporating our data-driven expressions into non-linear optimization problems. Finally, numerical simulations demonstrate the effectiveness of the proposed approach.
 
\smallskip
\noindent
\textbf{Paper organization.}
The paper is organized as follows. Section \ref{sec:prelim} introduces the problem setting, together with some classical results on pole placement and eigenstructure assignment. Section \ref{sec : closed form} presents our main expressions for data-driven pole placement and eigenstructure assignment. Section \ref{sec : optim} discusses strategies for the design of sparse controllers and validates them through numerical examples. The concluding remarks are left to Section \ref{sec : conclusions}.

\smallskip
\noindent
\textbf{Notation.}
Let $\real$ ($\mathbb N$, $\complex$) and
$\real^+$ ($\mathbb N^+$) denote the set of real (integer, complex) and strictly positive real (integer) numbers, 
respectively. Given a matrix
$A \in \real^{n \times m}$, $\Rank(A)$, 
$\Ker(A)$, and  $A^\top$ denote the rank, 
the kernel, and the transpose of $A$. $I_n$ and $0_{n,m}$ stand for the $n\times n$ identity matrix and $n\times m$ zero matrix, respectively (subscripts will be omitted when clear from the context). 
The $2$-norm of matrix $A$ is $\norm{A}$,
the Kronecker product between matrices $A$ and
$B$ is denoted by $A \otimes B$, and the Hadamard (elementwise) product by $A \circ B$. We let $\vect(\cdot) : \real^{n \times m} \rightarrow \real^{nm}$ denote the vectorization operator of a matrix. We let $\rho(A)$ denote the spectrum of matrix $A$, i.e., the set of eigenvalues of $A$. We let $\lambda^*$ denote the complex conjugate of $\lambda \in \complex$. For matrix $A \in \real^{m \times n}$ and $n$-dimensional subspace $\mathcal V$, we let $A \mathcal V = \{ Av_i, v_i \in \mathcal V\}$.

\section{Problem Setup and Preliminary Notions} \label{sec:prelim}
Consider a controllable discrete-time linear system
\begin{equation} \label{eq : sys}
	x(t+1) = Ax(t) + Bu(t),
\end{equation}
where $x(t) \in \real^n$ and $u(t) \in \real^m$ are the state and input vectors, respectively, at time $t \in \mathbb N$, and $A \in \real^{n \times n}$ and $B \in \real^{n \times m}$, and where we assume that $\Rank(B) = m$. In this paper we study the problem of computing a controller $K \in \real^{m \times n}$ which shapes the closed loop trajectory
\begin{equation} \label{eq : cl}
	x(t+1) = (A - BK)x(t)
\end{equation}
according to some design objectives. We assume that the model of the dynamical system \eqref{eq : sys}, i.e., matrices $A$ and $B$, is not available and, instead, we leverage a series of offline open-loop trajectories of \eqref{eq : sys}. In particular, we perform and collect data from $N \in \mathbb N^+$ experiments of length $T\in \mathbb N^+$, where $x^i(0)$, $\xbf^i_T = \vect{(x^i(1), \dots, x^i(T) )}$ and $\ubf^i_T = \vect(u^i(0), \dots, u^i(T-1))$ are the initial state, the state trajectory and the input trajectory, respectively, recorded for \eqref{eq : sys} during experiment $i \in \until{N}$. The dataset is available through matrices
\begin{subequations}\label{eq : data}
\begin{align} 
	X_0 &= 
	\begin{bmatrix}
		x^1(0) & x^2(0) & \dots & x^N(0)
	\end{bmatrix}, \\
	X &= 
	\begin{bmatrix}
		\xbf^1_T & \xbf^2_T & \dots & \xbf^N_T
	\end{bmatrix}, \text{ and} \\
	U &= 
	\begin{bmatrix}
		\ubf^1_T & \ubf^2_T & \dots & \ubf^N_T
	\end{bmatrix}.
\end{align}
\end{subequations}
The following Assumption and Lemma enable us to write any trajectory of \eqref{eq : sys} in terms of data collected as in \eqref{eq : data}. 
\begin{assumption}{\emph{\bfseries (Persistency of excitation)}} \label{ass : pe}
	Data matrices $X_0$ and $U$ in \eqref{eq : data} satisfy 
	\begin{equation*}
		\Rank\left(
		\begin{bmatrix}
			X_0 \\
			U
		\end{bmatrix}
		\right) = mT + n.
	\end{equation*}
	\oprocend
\end{assumption}
\begin{lemma}{\emph{\bfseries (Data-driven trajectories of \eqref{eq : sys} \cite{FC-FP:22})}}\label{lemma: trajectory combinations}
  Let \eqref{eq : data} be the dataset generated by 
  \eqref{eq : sys}, and let $\bar \xbf_T$ be any
  state trajectory of \eqref{eq : sys} generated with some
  initial condition $\bar x_0$ and control $\bar \ubf_T$. Then, there always exist vectors $\alpha$ and $\beta$, of appropriate dimension, such that
  \begin{equation*}\label{eq : x and y bar}
    \bar \xbf_T
    =
    \begin{bmatrix}
      XK_U & XK_0
    \end{bmatrix}
    \begin{bmatrix}
      \alpha\\ 
      \beta
    \end{bmatrix}
    ,
  \end{equation*}
  where $K_U = \Basis(\Ker(U))$ and $K_0 = \Basis(\Ker(X_0))$. 
  Moreover, $\bar x_0 = X_0 K_U \alpha$ and $\bar \ubf_T = U K_0 \beta$. \oprocend
\end{lemma}

Assumption \ref{ass : pe} is typical in data-driven studies and leverages the linearity of \eqref{eq : sys} to ensure that the collected dataset \eqref{eq : data} is sufficiently informative, i.e., that any state trajectory $\bar \xbf_T$ of \eqref{eq : sys} can be expressed as a linear combination of the recorded state trajectories $X$. Lemma \ref{lemma: trajectory combinations} goes a step further by decomposing $\bar \xbf_T$ into its free response ($X K_U \alpha$) and forced response ($X K_0 \beta$), and by expressing them as a function of data, for given initial condition $\bar x_0$ and input $\bar \ubf_T$.

In the following we let $\mathcal L = \fromto{\lambda_1}{\lambda_n}$ be the set of desired closed-loop eigenvalues and $\mathcal V = \fromto{\upsilon_1}{\upsilon_n}$ be the set of desired closed-loop eigenvectros, with $\lambda_i\in\real$ and $\upsilon_i \in \real^n$, for all $i \in \until{n}$.\footnote{Although both left and right eigenvectors can be considered, in this paper we shall refer to the right eigenvectors.} 

\begin{assumption} {\emph{\bfseries (Properties of closed-loop eigenvalues and eigenvectors)}} \label{ass : real eigen}
	We assume that the set of desired eigenvalues $\mathcal L$ of $A-BK$ is closed under complex conjugation, and that, for each eigenvalue, the geometric multiplicity matches the algebraic multiplicity. Eigenvectors corresponding to complex conjugate eigenvalues are complex conjugate. \oprocend
\end{assumption}

We leave the problem of generalizing the results of this paper to eigenvalues with different geometric and algebraic multiplicity as the focus for future research, and note that the condition on the multiplicity of the eigenvalues in Assumption \ref{ass : real eigen} is met when the elements of $\mathcal L$ are distinct.

An important limitation of the eigenstructure assignment problem is that it does not allow for arbitrary selection of eigenvalue/eigenvector pairs through feedback gain $K$. Instead, the choice of each eigenvector is constrained to a specific subspace within the system's state space. This restriction is formally expressed in the following theorem.
 
\begin{theorem}{\emph{\bfseries (Feasibility of eigenstructure assignment \cite{GPL-RJP:98})}} \label{thm : freedom}
	Using the control law $u(t) = -Kx(t)$ in \eqref{eq : sys} $n$ eigenvalues of $A-BK$ may be assigned and $m$ entries of each corresponding eigenvector can be chosen freely. \oprocend
\end{theorem}

Theorem \ref{thm : freedom} ensures that $\mathcal L = \fromto{\lambda_1}{\lambda_n}$ eigenvalues can be freely assigned, but restricts the choice of each eigenvector $\upsilon_i \in \real^n$ in $\mathcal V = \fromto{\upsilon_1}{\upsilon_n}$, to the subspace $\mathcal P_i \subseteq \real^n$, termed the \emph{allowable eigenvector subspace} \cite{GPL-RJP:98}. We recall a second fundamental result which ensures the existence and uniqueness of $K$ under certain conditions.

\begin{theorem}{\emph{\bfseries (Uniqueness of feedback for eigenstructure assignment	 \cite{BCM:76})}} \label{thm : unique K}
	Let $\mathcal L = \fromto{\lambda_1}{\lambda_n}$
  and $\mathcal V=\fromto{\upsilon_1}{\upsilon_n}$, with
  $\upsilon_i \in \mathcal P_i$, be the set of desired closed loop
  eigenvalues and eigenvectors, with  $\Rank(\begin{bmatrix}
		\upsilon_1 & \cdots & \upsilon_n
	\end{bmatrix}) = n$. Let Assumption \ref{ass : real eigen} hold, and let $\Rank(B) = m$. Then, the matrix $K$ such that $A-BK$ has eigenvalues in $\mathcal L$ and eigenvectors in $\mathcal V$ exists and is unique. \oprocend
\end{theorem}

In the pole placement problem, the set of desired closed-loop
eigenvectors is not specified. This leaves some degrees of freedom in
the selection of the columns of $\mathcal V$, as long as
$\upsilon_i \in \mathcal P_i$ (cf. Theorem \ref{thm :
  allowable eigenvectors}).  In general, the matrix $K$ in the pole
placement problem is, therefore, not unique. In the next section we
give a data-driven expression for $\mathcal P_i$, as well as a
data-driven characterization of the set of matrices $K$ such that
$\rho(A-BK) = \mathcal L$. Further, we show that the unique $K$ can be
found as a closed-form function of the data for the eigenstructure
assignment problem. In Section \ref{sec : optim} we leverage the
flexibility on $K$ for the pole placement problem, by introducing
additional design goals, i.e., enforcing sparsity constraints on $K$.

%

\section{Data-driven Pole Placement and Eigenstructure Assignment} \label{sec : closed form}

We begin with a data-driven expression to compute $\mathcal P_i$, the allowable eigenvector subspace associated with eigenvalue $\lambda_i$ for system \eqref{eq : sys}. That is, we wish to find the subspace $\mathcal P_i$ such that $(A-BK) \upsilon_i = \lambda_i \upsilon_i$, for all $\upsilon_i \in \mathcal P_i$, and some $K$. We remark that $\mathcal P_i$ is independent of $K$ \cite{GPL-RJP:98}, and in this paper we compute $\mathcal P_i$ without any explicit knowledge of system matrices $A$ and $B$, but leveraging only offline data \eqref{eq : data}. Throughout the paper, given $\lambda_i \in \mathcal L$, we define 
\begin{equation} \label{eq : Lambda i}
	\Lambda_i = \begin{bmatrix}
			I &
			\lambda_i I &
			\lambda_i^2 I &
			\cdots &
			\lambda_i^{T-1} I &
			\lambda_i^T I
		\end{bmatrix}^\top \in \real^{n(T+1) \times n},
\end{equation}
and matrices $Z = \begin{bmatrix}
	I_{nT} & 0_{nT \times n}
\end{bmatrix}$ (i.e., the matrix that extracts the first $nT$ rows from $\Lambda_i$) and $W = \begin{bmatrix}
	0_{nT \times n} & I_{nT}
\end{bmatrix}$ (i.e., the matrix that extracts the last $nT$ rows from $\Lambda_i$).

\begin{theorem}{\emph{\bfseries (Data-driven allowable eigenvector
      subspace)}} \label{thm : allowable eigenvectors} Let $\lambda_i$
  be a desired closed loop eigenvalue. The eigenvector associated with
  $\lambda_i$ belongs to the following~subspace:
	\begin{equation} \label{eq : alpha eigenvector}
		\mathcal P_{i} = X_0 K_U
		\begin{bmatrix}
			I & 0	
		\end{bmatrix}
		\Ker\left(
		\begin{bmatrix}
			X K_U - W \Lambda_i X_0 K_U & X K_0
		\end{bmatrix}\right).
	\end{equation}
\end{theorem}
\begin{proof}
First, we notice that if $\lambda_i$ and $\upsilon_i$ are an eigenvalue and the corresponding eigenvector of $A-BK$, then the following must hold
\begin{equation}
	(A-BK)\upsilon_i = \lambda_i \upsilon_i.		
\end{equation} 	
Further, only a trajectory $\xbf_T$ starting in $\upsilon_i$ will always remain in $\upsilon_i$ when evolving according to \eqref{eq : cl}. That is, if and only if $x(0) = \upsilon_i \in \mathcal P_i$, then $x(t) \in \Image(\upsilon_i)$ for all times $t$, and $x(t+1) = \lambda_i x(t)$. For a trajectory of length $T$, this can be written as
\begin{equation} \label{eq : supp 1.1}
	\xbf_T	
	=
	\begin{bmatrix}
		\lambda_i I \\
		\lambda_i^2 I \\
		\vdots \\
		\lambda_i^T I
	\end{bmatrix} x(0) 
	= 
	\begin{bmatrix}
		X K_U & X K_0
	\end{bmatrix}
	\begin{bmatrix}
		\alpha \\
		\beta
	\end{bmatrix}
\end{equation}
if and only if $x(0) \in \mathcal P_i$. From the rightmost equality in \eqref{eq : supp 1.1} and by noticing that $x(0) = X_0 K_U \alpha$ (cf. Lemma \ref{lemma: trajectory combinations}) one can conclude that $x(0) \in \mathcal P_i$ if and only if $\alpha$ and $\beta$ verify  
\begin{equation} \label{eq : supp 1.2}
	\begin{bmatrix}
		\alpha \\
		\beta 
	\end{bmatrix}
	\in 
	\Ker \left(
	\begin{bmatrix}
		X K_U - W \Lambda_i X_0 K_U & X K_0
	\end{bmatrix}
	\right).
\end{equation}
Extracting $\alpha$ from \eqref{eq : supp 1.2} and recalling that 
\begin{equation} \label{eq : supp 1.3}
	x(0) = \upsilon_i = X_0 K_U \alpha,	
\end{equation}
concludes the proof.
\end{proof}

Through Theorem \ref{thm : allowable eigenvectors} one can write the allowable eigenvector subspace associated to eigenvalue $\lambda_i$ for a closed loop dynamics \eqref{eq : cl} as a function of the open loop data \eqref{eq : data}.
However, Theorem \ref{thm : allowable eigenvectors} cannot be used to compute a static feedback controller $K$. In fact, simply imposing $\ubf_T = U K_0 \beta$ with $\beta $ as in \eqref{eq : supp 1.2} might result in an input generated by a non-static feedback. Next, we give a condition that restricts the choice of $\beta$ so that $\ubf_T$ is the result of a static feedback of the state, i.e. $u(t) = -Kx(t)$. Specifically, in Theorem \ref{thm : compatible feedbacks} we characterize the set of all static feedback $K$ that precisely place the eigenvalues of $A-BK$ to the desired set $\mathcal L$. 

\begin{theorem}{\emph{\bfseries (Data-driven pole
      placement)}} \label{thm : compatible feedbacks} Let
  $\mathcal L = \fromto{\lambda_1}{\lambda_n}$ be the desired closed
  loop eigenvalues. Then, $\rho(A-BK)=\mathcal L$ if and only if
  $K \in \mc K$, where
  \begin{equation} \label{eq : set K}
    \mathcal K = \left\{K : \bigcap_{i = 1}^{n}
      \Ker \left(
        \begin{bmatrix}
          (I \otimes K) Z \Lambda_i X_0 K_U & U K_0
        \end{bmatrix} \right) \neq 0 \right\}.
  \end{equation}
\end{theorem}
\begin{proof}
	Let $\lambda_i$ be a desired closed loop eigenvalue and let $x(0) \in \mathcal P_i$. Then $x(t+1) = \lambda_i x(t) = \lambda_i^{t+1} x(0)$. From $u(t) = -K x(t) = -K \lambda_i^t x(0)$ we write 
	\begin{equation} \label{eq : supp 3.1}
		\ubf_T =
		-\begin{bmatrix}
			K  \\
			& K \\
			& & K  \\
			& & & \ddots \\
			& & & & K
		\end{bmatrix}
		\begin{bmatrix}
			I \\ 
			\lambda_i I \\
			\lambda_i^2 I \\
			\vdots \\
			\lambda_i^{T-1} I
		\end{bmatrix}
		x(0).
	\end{equation}
	We recall that $\ubf_T = U K_0 \beta$ and $x(0) = X_0 K_U \alpha$ in \eqref{eq : supp 3.1} (cf. Lemma \ref{lemma: trajectory combinations}) and therefore we can write  
	\begin{equation} \label{eq : supp 3.2}
		U K_0 \beta = (I_T \otimes -K) Z \Lambda_i X_0 K_U \alpha,
	\end{equation}
	for all $i \in \until{n}$. Equation \eqref{eq : supp 3.2} needs to be verified for every $\Lambda_i$ and therefore the vectors $\alpha$ and $\beta$ need to satisfy
	\begin{equation} \label{eq : beta feedback}
		\begin{bmatrix}
			\alpha \\
			\beta
		\end{bmatrix}
		\in
		\bigcap_{i = 1}^{n}
		\Ker \left(
		\begin{bmatrix}
			(I \otimes K) Z \Lambda_i X_0 K_U & U K_0
		\end{bmatrix}
		\right).
	\end{equation}
That is, $K$ is a static feedback (cf. \eqref{eq : supp 3.1}) such that $\rho(A-BK) = \mathcal L$ if and only if $\alpha$ and $\beta$ in \eqref{eq : beta feedback} exist. From this conclusion, the condition \eqref{eq : set K} on $\mathcal K$ is directly derived. 
\end{proof}

Through Theorem \ref{thm : compatible feedbacks} we can characterize the set of feedback gains such that $\rho(A-BK)=\mathcal L$, that is, \emph{all} the feedback gains which solve the pole placement problem for a given set $\mathcal L$. This condition will be used in Section \ref{sec : optim} with the aim of extracting matrices $K$ from $\mathcal K$ which satisfy some desired sparsity pattern.

We conclude this section by leveraging Theorem \ref{thm : allowable eigenvectors} and \ref{thm : compatible feedbacks} to find a closed-from expression for the eigenstructure assignment problem, i.e., when both $\mathcal L$ and $\mathcal V$ are given. For simplicity, and without affecting the generality of the approach, we limit the data collection phase in \eqref{eq : data} to $T=1$.
 
\begin{theorem}{\emph{\bfseries (Closed-form expression of the
      feedback gain for eigenstructure assignment)}} \label{thm :
    closed form K} Let $\mathcal L = \fromto{\lambda_1}{\lambda_n}$
  and $\mathcal V=\fromto{\upsilon_1}{\upsilon_n}$, with
  $\upsilon_i \in \mathcal P_i$, be the set of desired closed loop
  eigenvalues and eigenvectors. Let
  \begin{equation} \label{eq : alpha beta cf}
    \begin{bmatrix}
      \alpha_i \\
      \beta_i
    \end{bmatrix}
    =
    \Basis\left(\Ker\left(
        \begin{bmatrix}
          X K_U - \lambda_i X_0 K_U & X K_0
        \end{bmatrix}\right) \right) \gamma_i,
  \end{equation}
  where $X$, $X_0$, and $U$ are as in \eqref{eq : data} with $T = 1$,
  and $\gamma_i$ satisfies
  $\Basis(\mathcal P_i)\gamma_i = \upsilon_i$. The closed-loop matrix
  $A-BK$, with
  \begin{equation} \label{eq : closed form K}
    K = - U K_0
    \begin{bmatrix}
      \beta_1 & \cdots & \beta_n 
    \end{bmatrix}
    \left(
      X_0K_U
      \begin{bmatrix}
        \alpha_1 & \cdots & \alpha_n
      \end{bmatrix}
    \right)^{-1},
  \end{equation}
  has eigenvalues $\mathcal L$ and eigenvectors $\mathcal V$. \oprocend
\end{theorem}
\begin{proof}
 From condition \eqref{eq : beta feedback}, we seek the $K$ such that 
	\begin{equation} \label{eq : supp 4.1}
	\begin{bmatrix}
		\alpha_i \\
		\beta_i
	\end{bmatrix}
	\in 
	\Ker\left(
	\begin{bmatrix}
		K X_0 U & U K_0
	\end{bmatrix} 
	\right),	~ \forall i = \until{n},
	\end{equation}
	with $\alpha_i, \beta_i$ defined in \eqref{eq : alpha beta cf}, and where $(I_T \otimes K) Z \Lambda_i =  K$ from $T=1$. We can write condition \eqref{eq : supp 4.1} on $K$ as
	\begin{align}
		\begin{bmatrix}
		 K X_0 K_U & U K_0
		\end{bmatrix}
		\begin{bmatrix}
			\alpha_1 & \alpha_2 & \cdots & \alpha_n \\
			\beta_1 & \beta_2 & \cdots & \beta_n
		\end{bmatrix}
		= 0,
	\end{align}
	from which we find that 
	$$K X_0 K_U 
	\begin{bmatrix}
		\alpha_1 & \cdots & \alpha_n
	\end{bmatrix}
	= 
	-U K_0 
	\begin{bmatrix}
		\beta_1 & \cdots & \beta_n
	\end{bmatrix}.
	$$
	We notice that $
		X_0K_U
		\begin{bmatrix}
			\alpha_1 & \cdots & \alpha_n
		\end{bmatrix}$ is invertible since both $X_0 K_U$ and $\begin{bmatrix}
			\alpha_1 & \cdots & \alpha_n
		\end{bmatrix}$ are square matrices with full rank.\footnote{The fact that $X_0 K_U$ is full rank is a direct consequence of Assumption \ref{ass : pe}, see also \cite{FC-FP:22}. The fact that $\Gamma = \begin{bmatrix}
			\alpha_1 & \cdots & \alpha_n
		\end{bmatrix}$ is full rank can be proved by contradiction. Assume, without loss of generality, that $\{\delta_1 , \cdots , \delta_{n-1}\}$, $\delta_i \in \real$, exist such that $\alpha_n = \sum_{k = 1}^{n-1}\delta_k\alpha_k$, i.e., $\alpha_n$ is a linear combination of the remaining columns of $\Gamma$, and therefore $\Gamma$ is singular. Then, from \eqref{eq : supp 1.3}, $\upsilon_n = X_0 K_U \alpha_n = X_0 K_U \sum_{k = 1}^{n-1}\delta_k\alpha_k = \sum_{k = 1}^{n-1} \delta_k \upsilon_k$. This would imply that the elements in $\mathcal V$ are not linearly independent, contradicting Assumption \ref{ass : real eigen}.} This ensures the existence and uniqueness of $K$ and concludes the proof.
\end{proof}

Thanks to Theorem \ref{thm : closed form K} a closed-form solution for the eigenstructure assignment problem is found. When needed, Theorem \ref{thm : closed form K} can be also used to find a solution for the pole placement problem by simply selecting any arbitrary $\mathcal V$ such that $\upsilon_i \in \mathcal P_i$, by leveraging Theorem \ref{thm : allowable eigenvectors}. As we have discussed, fixing $\mathcal L$ while leaving more freedom on the choice of eigenvectors $\mathcal V$ renders $K$ not unique. This allows for more interesting problems to be solved, for example, by imposing some sparsity constraints on matrix $K$. In the next section we explore strategies to leverage Theorem \ref{thm : allowable eigenvectors} and Theorem \ref{thm : compatible feedbacks} to find a sparse $K \in \mathcal K$. These solutions are not closed-form but rather based on the solution of bilinear optimization programs.

\section{Data-driven Pole Placement with Sparse Feedback
  Matrices} \label{sec : optim} Consider the pole placement problem
with desired closed loop eigenvalues $\mathcal L$. As previously
discussed, each eigenvector $\upsilon_i$ corresponding to eigenvalue
$\lambda_i$ must belong to a subspace $\mathcal P_i$. Let $\mathcal K$
in \eqref{eq : set K} be the set containing all matrices $K$ such that
$\rho(A-BK) = \mathcal L$. Then, we can look for a pair of $K$ and
$\mathcal V=\fromto{\upsilon_1}{\upsilon_n}$ that satisfy
\begin{argmini}|l|[0]
    {K, \mathcal V}{ f_1(K)}{}
    {\label{eq : min general}}
    \addConstraint{ 
    K \in \mathcal K
    }{}
    \addConstraint{\upsilon_i \in \mathcal  P_i, \quad \forall i \in \until{n}}{}
    \addConstraint{f_2(K) = 0,}{}
\end{argmini}
where $f_1(K) : \real^{m\times n} \rightarrow \real$ and $f_2(K) : \real^{m\times n} \rightarrow \real^q$ are some functions of $K$. We notice that \eqref{eq : min general} is a bilinear optimization problem in the variables $K$ and $\mathcal V$, which follows from the definition of $\mathcal K$ in \eqref{eq : set K}. 
Different choices of $f_1(K)$ and $f_2(K)$ lead to the solution of different problems, as will be detailed next.

%

\subsection{Data-driven Minimum-gain Pole Placement with Sparse Static Feedback}
Problem \eqref{eq : min general} can be cast as a data-driven optimization problem thanks to the results of Section \ref{sec : closed form}. In particular, the condition on $\upsilon_i \in \mathcal P_i$ can be imposed through Theorem \ref{thm : allowable eigenvectors}, while $\mathcal K$ is characterized in Theorem \ref{thm : compatible feedbacks}. 
We now propose an optimization-based strategy to compute a static feedback $K$ with sparsity constraints, directly from data. 

Let $S \in \{0,1\}^{m\times n}$ be the binary matrix that specifies the sparsity structure of feedback $K$. That is, we wish to find $K$ such that
\begin{equation}
	K_{i j}= \begin{cases}0 & \text { if } S_{i j}=1, \\ \star & \text { if } S_{i j}=0.\end{cases}
\end{equation}
Now, consider the following optimization problem to find a $K \in \mathcal K$ which has sparsity constraints as specified by $S$
\begin{argmini}|l|[0]
    {\substack{K \\ \fromto{\gamma_1}{\gamma_n}}}{\frac{1}{2} \norm{K}_F^2}{}
    {\label{eq : min mgpp}}
    \addConstraint{ 
    \norm{
    \begin{bmatrix}
    	(I_T \otimes K) Z \Lambda_i X_0 K_U & U K_0
    \end{bmatrix}w_i
    } = 0}{}
    \addConstraint{w_i = \Ker\left(
    \begin{bmatrix}
    	(X - W \Lambda_i X_0)K_U & X K_0
    \end{bmatrix}
    \right) \gamma_i}{}
     \addConstraint{S \circ K = 0_{m \times n}.}{}
\end{argmini}
The above is a data-driven implementation of \eqref{eq : min general}, with $f_1(K) = \norm{K}_F^2$ and $f_2(K) = S \circ K$. Minimizing the norm of the gain in $f_1(K)$ reduces the overall control effort, while $f_2(K)$ imposes the desired sparsity pattern on $K$. We notice that $\upsilon_i$ depends on the choice of $\gamma_i$, since $\upsilon_i = X_0 K_U \begin{bmatrix} I & 0 \end{bmatrix} w_i$, a direct consequence of \eqref{eq : alpha eigenvector}. 

\begin{remark}{\emph{\bfseries (Feasibility of \eqref{eq : min mgpp})}}
	There is no known procedure to determine the feasibility of \eqref{eq : min mgpp}, i.e., if $K \in \mathcal K$  exists such that $S \circ K =0$. In general, assessing the existence of a sparse static feedback $K$ is an NP-hard problem even when $(A,B)$ are known \cite{VK-FP:18a}. Therefore, in the following, we shall assume the feasibility of \eqref{eq : min mgpp}. We refer the interested reader to \cite{VK-FP:18a} for a detailed analytical characterization of the locally optimal solution of \eqref{eq : min general} in terms of the eigenvector matrices of the closed-loop system.	\oprocend
\end{remark}

\begin{remark}{\emph{\bfseries (Fixed modes of $(A-BK$) \cite{MES-DDS:81})}}
	We recall that the fixed modes of $(A,B)$ with respect to the sparsity constraints $S$ are the eigenvalues of $A$ that cannot be changed using a sparse state feedback. When the fixed modes of $(A,B)$ do not belong to $\mathcal L$, problem \eqref{eq : min mgpp} becomes unfeasible and a different sparsity constraint $S$ must be selected. \oprocend
\end{remark}
 
We now discuss a numerical implementation of \eqref{eq : min mgpp}.
 
\begin{example}{\emph{\bfseries (Data-driven sparse feedback)}} \label{ex : mgpp}
	We consider the discretized version of a batch reactor system \cite{GCW-HY:01}
	(with sampling time of $0.1\text{s}$), with
	\begin{align*}
	\scriptsize		
	A = \begin{bmatrix}
			1.178 & 0.001  & 0.511  & -0.403 \\
	 	   -0.051 & 0.661  & -0.011 & 0.061 \\
	  		0.076 & 0.335  & 0.560  & 0.382 \\
	  		0     & 0.335  & 0.089  & 0.849
		\end{bmatrix}\!, \hspace*{-1em}
		\quad
		B = \begin{bmatrix}
			0.004 & -0.087 \\
	 		0.467 &  0.001 \\
	 		0.213 & -0.235 \\
	 		0.213 & -0.016
		\end{bmatrix}\!,
	\end{align*}
	which is open-loop unstable with $\sigma(A) = \{ 1.2200,
    1.0049,
    0.4206,
    0.6025\}$. We set $T=10$ and collect a series
	of $N = n + mT = 24$ experiments \eqref{eq : data}
	satisfying Assumption \ref{ass : pe}. We select the desired closed-loop eigenvalues $\mathcal L = \{-0.3, 0.2, 0.5, 0.7 \}$ and the sparsity pattern
	\begin{align*}
		S &= \begin{bmatrix}
			1 & 0 & 0 & 0 \\
			0 & 0 & 1 & 0
		\end{bmatrix}.
	\end{align*}
	Problem \eqref{eq : min mgpp} is implemented in \texttt{MATLAB} and solved using the \texttt{fmincon} routine. A stabilizing $K$, satisfying the conditions required by $\mathcal L$ and $S$ is found, with
	\begin{equation}
		K = 
		\begin{bmatrix*}
			  0.0000  &  2.7633 &   2.7324  &  0.4122 \\
  			 -2.3621  &  1.2654  & -0.0000  &  1.1906
		\end{bmatrix*},
	\end{equation}
	and
	\begin{equation}
		V = 
		\begin{bmatrix*}		
    0.0475  & -0.1938 &  -0.2007 &  -0.5204 \\
    0.9606  & -0.8211 &  -0.6873 &   0.3220 \\
   -0.2581  &  0.5266 &   0.5699 &  -0.2354 \\
    0.0914  &  0.1046 &   0.4032 &  -0.7550
		\end{bmatrix*},
	\end{equation}
	where $V = \begin{bmatrix}
		\upsilon_1 & \cdots & \upsilon_n
	\end{bmatrix}$.
	We remark that $K$ and $V$ are not unique, in general. 
	\hfill $\square$
\end{example}

\subsection{Data-driven Maximally Sparse Feedback}
As a second example of the application of the results of Section \ref{sec : closed form} to the data-driven design of sparse controllers, we consider the problem of finding the maximally sparse feedback $K$. In this scenario we are not seeking for a feedback with a defined sparsity pattern, as specified in $S$, but rather the one which has the most number of entries at zero. This is done by removing the specification on the sparsity pattern and by minimizing $\sum_{ij} | K_{ij}|$ which is often used as a proxy for the $L0$-norm of a matrix. In this case, problem \eqref{eq : min general} can be written as 
\begin{argmini}|l|[0]
    {\substack{K \\ \fromto{\gamma_1}{\gamma_n} 
    }}{\sum_{ij} |K_{ij}|}{}
    {\label{eq : min max sparse}}
    \addConstraint{ 
    {
    \begin{bmatrix}
    	(I_T \otimes K) Z \Lambda_i X_0 K_U & U K_0
    \end{bmatrix}w_i
    } = 0}{}
    \addConstraint{w_i = \Ker\left(
    \begin{bmatrix}
    	(X - W \Lambda_i X_0)K_U & X K_0
    \end{bmatrix}
    \right) \gamma_i.}{}
\end{argmini}

This is, again, a bilinear optimization problem. We now show a numerical implementation of this approach. 

\begin{example} {\emph{\bfseries (Data-driven maximally sparse feedback) }}\label{ex : max sparsity}
	Consider the same problem settings as in Example \ref{ex : mgpp}. We let $\mathcal L = \{-0.3, 0.2, 0.5, 0.7 \}$ be sets of the desired closed-loop eigenvalues. By running \eqref{eq : min max sparse} we find
		\begin{equation*}
			K = 
		\begin{bmatrix}	  
    0.0000 &   1.6901 &   0.0000 &   4.4741 \\
   -1.9515 &   0.0000 &  -1.0042 &   0.0000
		\end{bmatrix}
		\end{equation*}
		and
		\begin{equation*}
			V = \begin{bmatrix}
    0.7460 &   0.5153 &   0.0752 &  -0.0053 \\
    0.1892 &   0.4018 &   0.9439 &   0.9901 \\
   -0.6318 &  -0.7451 &  -0.2829 &   0.1127 \\
   -0.0922 &  -0.1332 &  -0.1532 &   0.0834
			\end{bmatrix}.
		\end{equation*}
		Despite the non-convexity of \eqref{eq : min max sparse} we obtain a sparse controller with a total of $4$ entries at zero. \oprocend
\end{example}

\section{Conclusions} \label{sec : conclusions}
In this paper we consider a data-driven strategy for the design of, possibly sparse, feedback gains for pole placement and eigenstructure assignment. Given a set of desired closed-loop eigenvalues $\mathcal L$, we characterize the allowable eigenvector subspaces of the dynamical system described by unknown $(A,B)$, as well as the set $\mathcal K$ of feedback gains such that $\rho(A-BK)=\mathcal L$ for all $K \in \mathcal K$. For the eigenstructure assignment problem, we give a closed-form data-driven expression for the gain $K$ which assigns the desired closed-loop eigenvalues $\mathcal L$ together with the associated desired eigenvectors $\mathcal V$. Further, we discuss optimization-based strategies to find a sparse gain $K \in \mathcal K$ when a desired sparsity structure needs to be imposed on $K$, or when the overall sparsity of $K$ (i.e., the number of zero elements of $K$) needs to be maximized. Numerical simulations complement our analysis. Future work includes a characterization of the performance of these tools when data is collected with noise, together with a comparison with model-based methods based on system identification from data. 

\bibliographystyle{ieeetr}
\bibliography{alias,FP,Main,New}

\begin{thebibliography}{10}

\bibitem{GB-DSB-FP:20}
G.~Baggio, D.~S. Bassett, and F.~Pasqualetti, ``Data-driven control of complex
  networks,'' {\em Nature Communications}, vol.~12, no.~1429, 2021.

\bibitem{MP-SS-MT-CC-AK-RS-JN:18}
M.~Pfeiffer, S.~Shukla, M.~Turchetta, C.~Cadena, A.~Krause, R.~Siegwart, and
  J.~Nieto, ``Reinforced imitation: Sample efficient deep reinforcement
  learning for mapless navigation by leveraging prior demonstrations,'' {\em
  IEEE Robotics and Automation Letters}, vol.~3, no.~4, pp.~4423--4430, 2018.

\bibitem{XX:22}
X.~Xiao, T.~Zhang, K.~Choromanski, E.~Lee, A.~Francis, J.~Varley, S.~Tu,
  S.~Singh, P.~Xu, F.~Xia, S.~M. Persson, D.~Kalashnikov, L.~Takayama,
  R.~Frostig, J.~Tan, C.~Parada, and V.~Sindhwani, ``Learning model predictive
  controllers with real-time attention for real-world navigation,'' in {\em
  Proceedings of The 6th Conference on Robot Learning}, pp.~1708--1721, 2023.

\bibitem{SM-RRH:22}
S.~Mukherjee and R.~R. Hossain, ``Data-driven pole placement in {LMI} regions
  with robustness guarantees,'' in {\em 2022 IEEE 61st Conference on Decision
  and Control (CDC)}, pp.~4010--4015, IEEE, 2022.

\bibitem{GB:23}
G.~Bianchin, ``Data-driven exact pole placement for linear systems,'' {\em
  arXiv preprint arXiv:2303.11469}, 2023.

\bibitem{MI-CS:90}
M.~Innocenti and C.~Stanziola, ``Performance—robustness trade off of
  eigenstructure assignment applied to rotorcraft,'' {\em The Aeronautical
  Journal}, vol.~94, no.~934, pp.~124--131, 1990.

\bibitem{ANA-EYS-JCC:83}
A.~N. Andry, E.~Y. Shapiro, and J.~C. Chung, ``Eigenstructure assignment for
  linear systems,'' {\em IEEE Transactions on Aerospace and Electronic
  Systems}, no.~5, pp.~711--729, 1983.

\bibitem{GPL-RJP:98}
G.~P. Liu and R.~J. Patton, {\em Eigenstructure assignment for control system
  design}.
\newblock John Wiley \& Sons, Inc., 1998.

\bibitem{WMW:67}
W.~M. Wonham, ``On pole assignment in multi-input controllable linear
  systems,'' {\em IEEE Transactions on Automatic Control}, vol.~12, no.~6,
  pp.~660--665, 1967.

\bibitem{WMW-MAS:70}
W.~M. Wonham and A.~S. Morse, ``Decoupling and pole assignment in linear
  multivariable systems: a geometric approach,'' {\em SIAM Journal on Control},
  vol.~8, no.~1, pp.~1--18, 1970.

\bibitem{BCM:76}
B.~C. Moore, ``On the flexibility offered by state feedback in multivariable
  systems beyond closed loop eigenvalue assignment,'' {\em IEEE Transactions on
  Automatic Control}, vol.~21, no.~5, pp.~689--692, 1976.

\bibitem{GK-BCM:77}
G.~Klein and B.~C. Moore, ``Eigenvalue-generalized eigenvector assignment with
  state feedback,'' {\em IEEE Transactions on Automatic Control}, vol.~22,
  no.~1, pp.~140--141, 1977.

\bibitem{JK-NKN-PVD:85}
J.~Kautsky, N.~K. Nichols, and P.~{Van Dooren}, ``Robust pole assignment in
  linear state feedback,'' {\em International Journal of control}, vol.~41,
  no.~5, pp.~1129--1155, 1985.

\bibitem{FP-AF-LN:21}
F.~Padula, A.~Ferrante, and L.~Ntogramatzidis, ``Eigenstructure assignment in
  linear geometric control,'' {\em Automatica}, vol.~124, p.~109363, 2021.

\bibitem{JQT-MGT-NSF-SJE:22}
J.~Q. Teoh, M.~G. Tehrani, N.~S. Ferguson, and S.~J. Elliott, ``Eigenvalue
  sensitivity minimisation for robust pole placement by the receptance
  method,'' {\em Mechanical Systems and Signal Processing}, vol.~173,
  p.~108974, 2022.

\bibitem{FL-MF-MRJ:11al}
F.~Lin, M.~Fardad, and M.~R. Jovanovi\'c, ``Augmented {L}agrangian approach to
  design of structured optimal state feedback gains,'' {\em IEEE Transactions
  on Automatic Control}, vol.~56, no.~12, pp.~2923--2929, 2011.

\bibitem{GB-VK-FP:19}
G.~Baggio, V.~Katewa, and F.~Pasqualetti, ``Data-driven minimum-energy controls
  for linear systems,'' {\em IEEE Control Systems Letters}, vol.~3, no.~3,
  pp.~589--594, 2019.

\bibitem{JC-JL-FD:18}
J.~Coulson, J.~Lygeros, and F.~D{\"o}rfler, ``Data-enabled predictive control:
  In the shallows of the {D}ee{PC},'' in {\em {E}uropean {C}ontrol
  {C}onference}, (Naples, Italy), pp.~307--312, 2019.

\bibitem{CDP-PT:19}
C.~{De Persis} and P.~Tesi, ``Formulas for data-driven control: Stabilization,
  optimality and robustness,'' {\em IEEE Transactions on Automatic Control},
  vol.~65, no.~3, pp.~909--924, 2020.

\bibitem{JWP-JCW:97}
J.~W. Polderman and J.~C. Willems, {\em Introduction to mathematical systems
  theory: a behavioral approach}, vol.~26.
\newblock Springer Science \& Business Media, 1997.

\bibitem{JCW-PR-IM-BLMDM:05}
J.~C. Willems, P.~Rapisarda, I.~Markovsky, and B.~L.~M.~D. Moor, ``A note on
  persistency of excitation,'' {\em Systems \& Control Letters}, vol.~54,
  no.~4, pp.~325--329, 2005.

\bibitem{ST-RF-MS:22}
S.~Tu, R.~Frostig, and M.~Soltanolkotabi, ``Learning from many trajectories,''
  {\em arXiv preprint arXiv:2203.17193}, 2022.

\bibitem{VK-FP:18a}
V.~Katewa and F.~Pasqualetti, ``Minimum-gain pole placement with sparse static
  feedback,'' {\em IEEE Transactions on Automatic Control}, vol.~66, no.~8,
  pp.~1558--2523, 2021.

\bibitem{JE-JC:22}
J.~Eising and J.~Cort{\'e}s, ``Informativity for centralized design of
  distributed controllers for networked systems,'' in {\em {E}uropean {C}ontrol
  {C}onference}, pp.~681--686, 2022.

\bibitem{YSW-NM-JCD:19}
Y.-S. Wang, N.~Matni, and J.~C. Doyle, ``A system-level approach to controller
  synthesis,'' {\em IEEE Transactions on Automatic Control}, vol.~64, no.~10,
  pp.~4079--4093, 2019.

\bibitem{FC-FP:22}
F.~Celi and F.~Pasqualetti, ``Data-driven meets geometric control: Zero
  dynamics, subspace stabilization, and malicious attacks,'' {\em IEEE Control
  Systems Letters}, vol.~6, pp.~2569--2574, 2022.

\bibitem{MES-DDS:81}
M.~E. Sezer and D.~D. {\v{S}}iljak, ``Structurally fixed modes,'' {\em Systems
  \& Control Letters}, vol.~1, no.~1, pp.~60--64, 1981.

\end{thebibliography}

\end{document}